\documentclass[preprint,12pt]{elsarticle}

\usepackage{graphicx}
\usepackage{dcolumn}
\usepackage{bm}
\usepackage{amssymb}
\usepackage{hyperref}
\usepackage{color}

\begin{document}

\begin{frontmatter}

\title{Self-sustained activity of low firing rate in balanced networks}

\author{F. S. Borges$^1$, P. R. Protachevicz$^2$, R. F. O. Pena$^3$, E. L.
Lameu$^{4,5}$, G. S. V. Higa$^1$, A. H. Kihara$^1$, F. S. Matias$^{6,7}$, C. G.
Antonopoulos$^8$, R. de Pasquale$^9$, A. C. Roque$^3$, K. C. Iarosz$^{10}$, P.
Ji$^{11,12}$, A. M. Batista$^{2,13}$}
\address{$^1$Center for Mathematics, Computation, and Cognition, Federal
University of ABC, S\~ao Bernardo do Campo, SP, Brazil.}
\address{$^2$Graduate in Science Program - Physics, State University of Ponta
Grossa, PR, Brazil.}
\address{$^3$Laboratory of Neural Systems, Department of Physics, University of
S\~ao Paulo, Ribeir\~ao Preto, SP, Brazil.}
\address{$^4$National Institute for Space Research, S\~ao Jos\'e dos Campos,
SP, Brazil.}
\address{$^5$Department of Physics, Humboldt University, Berlin, Germany.}
\address{$^6$Institute of Physics, Federal University of Alagoas, Macei\'o, AL,
Brazil.}
\address{$^7$Cognitive Neuroimaging Unit, CEA DRF/I2BM, INSERM, Universit\'e
Paris-Sud, Universit\'e Paris-Saclay, F-91191 Gif/Yvette, France}
\address{$^8$Department of Mathematical Sciences, University of Essex, Wivenhoe
Park, UK.}
\address{$^9$Department of Physiology and Biophysics, ICB, University of S\~ao
Paulo, S\~ao Paulo, SP, Brazil.}
\address{$^{10}$Institute of Physics, University of S\~ao Paulo, S\~ao Paulo,
SP, Brazil.}
\address{$^{11}$Key Laboratory of Computational Neuroscience and Brain-Inspired
Intelligence (Fudan University), Ministry of Education, China.}
\address{$^{12}$Institute of Science and Technology for Brain-Inspired
Intelligence, Fudan  University,  Shanghai, China.}
\address{$^{13}$Department of Mathematics and Statistics, State University of
Ponta Grossa, Ponta Grossa, PR, Brazil.}

\cortext[cor]{Corresponding author: fernandodasilvaborges@gmail.com, protachevicz@gmail.com}

\date{\today}

\begin{abstract}
Self-sustained activity in the brain is observed in the absence of external stimuli 
and contributes to signal propagation, neural coding, and dynamic stability. It also 
plays an important role in cognitive processes. In this work, by means of studying 
intracellular recordings from CA1 neurons in rats and results from numerical simulations, 
we demonstrate that self-sustained activity presents high variability of patterns, such as 
low neural firing rates and activity in the form of small-bursts in distinct neurons. In 
our numerical simulations, we consider random networks composed of coupled, adaptive 
exponential integrate-and-fire neurons. The neural dynamics in the random networks simulates 
regular spiking (excitatory) and fast spiking (inhibitory) neurons. We show that both the 
connection probability and network size are fundamental properties that give rise to self-sustained 
activity in qualitative agreement with our experimental results. Finally, we provide a more 
detailed description of self-sustained activity in terms of lifetime distributions, synaptic 
conductances, and synaptic currents.
\end{abstract}


\end{frontmatter}

%%%%%%%%%%%%%%%%%%%%%%%%%%%%%%%%%%%%%%%
%%%%%%%%%%%%%%%%%%%%%%%%%%%%%%%%%%%%%%%

\section{Introduction}

Self-sustained activity (SSA), where neurons display persistent activity even
in the absence of external stimuli \cite{greicius03,fox05}, is observed in
diverse situations such as in \emph{in vitro} cortical cultures and in slice
preparations \cite{plenz96,sanchez00,shu03}, in \emph{in vivo} cortical
preparations \cite{timofeev00}, in slow-wave sleep \cite{steriade01}, in anesthesia
\cite{steriade93}, and in the resting state \cite{arieli95,mantini07}.
Electrophysiological recordings of SSA states show irregular neural spiking,
typically with low average frequencies of a few Hertz, obeying long-tailed
distributions \cite{hromadka08,oconnor10,buzsaki14}.

Many works have modelled neural networks with SSA by using random networks
composed of excitatory and inhibitory leaky, integrate-and-fire (LIF) neurons
with external background input
\cite{brunel00,vogels05a,parga07,kumar08,kriener14,ostojic14}. Other studies
have considered networks with non-random architectures, composed of LIF neurons
\cite{renart07,kaiser10,wang11,litwin12,potjans14} or nonlinear, two-dimensional
integrate-and-fire neuron models
\cite{compte06,izhikevich08,destexhe09,stratton10,tomov14,tomov16}. Both the
architecture and neuron types that comprise the network play an important
role in SSA states. Such states are generated and maintained by recurrent
interactions within networks of excitatory and inhibitory neurons. A stable SSA
state is related to strong recurrent excitation within the neural network,
which is restrained by inhibition to prevent runaway excitation. The balance
between excitation and inhibition in neural networks is considered critical
to maintain a SSA state \cite{shu03,brunel00,van96,haider06,taub13,nanou2018}. 
Kumar et al. \cite{kumar08} studied the effects of network size on SSA 
states. Barak and Tsodyks \cite{barak07} shown that combinations of synaptic
depression and facilitation result in different network dynamics. Triplett et
al. \cite{triplett18} analysed spontaneous activity in developing neural
networks and shown that networks of binary threshold neurons can form
structured patterns of neural activity.

In this work, we verify the existence of SSA in a random network 
composed of neurons with different intrinsic firing patterns (the so-called 
electrophysiological classes \cite{contreras04}). We consider the adaptive, 
exponential integrate-and-fire (AdEx) \cite{brette05} model with cortical neurons 
modelled as regular spiking (RS) cells with spike frequency adaptation and 
fast-spiking (FS) cells with a negligible level of adaptation. In such networks, 
depending on the excitatory synaptic strength, neurons can exhibit a transition 
from spiking to bursting synchronisation \cite{borges17,protachevicz18} and bistable 
firing patterns \cite{protachevicz19}. We find conditions in which unstructured, 
sparsely connected random networks of AdEx neurons
can display low frequency, self-sustained activity.

In particular, we show that not only the balance between excitation and 
inhibition but also the connection density and network size are both important factors 
for low frequency SSA. In balanced networks, high mean node-degree connectivity is necessary 
to give rise to low mean neural firing rates, and for such high values, large networks 
are necessary to support SSA states. In our computer simulations, we obtain qualitatively 
similar results to the ones we observed in our experimental recordings, where we use CA1 neurons 
whole-cell recordings in rats to demonstrate the possible variability of firing rate patterns 
observed in the brain. Our intracellular recordings show high variability of spontaneous activity 
patterns including low and irregular neural firing rates of approximately 1 Hz and spike-train 
power spectra with slow fluctuations \cite{litwin12}, and small-bursts activity in distinct 
recorded neurons. Interestingly, we show that these results can be reproduced qualitatively by 
our model with cortical neurons modelled as regular spiking (RS) cells with spike frequency 
adaptation and fast-spiking (FS) cells with a negligible level of adaptation.

The paper is organised as follows: in Sec. \ref{nnm_section}, we introduce the neural network 
model and in Sec. \ref{dmee_section}, the various quantities used to study self-sustained 
activity in our numerical simulations and the details of our electrophysiological experiments. 
In Sec. \ref{ssa_section}, we present our results on self-sustained activity in the numerical 
simulations and experimental data, and in the last section we discuss them and draw our conclusions.

%%%%%%%%%%%%%%%%%%%%%%%%%%%%%%%%%%%%%%%
%%%%%%%%%%%%%%%%%%%%%%%%%%%%%%%%%%%%%%%

\section{Neural network model}\label{nnm_section}

We start by building a random neural network of $N$ AdEx neurons by connecting them with probability 
$p$, where $p$ is the probability that any two neurons in the network are connected, excluding autapses 
(i.e. neurons connected to themselves, thus self-loops are not allowed). The $N$ neurons are split into 
excitatory and inhibitory neurons according to the ratio 4:1 (meaning that 80\% are excitatory and 20\% 
are inhibitory), following \cite{noback05}. The connection probability $p$ and the mean connection degree 
$K$ are associated by means of the relation
\begin{equation}\label{eqpK}
p=\frac{K}{N-1}.
\end{equation}

The dynamics of each AdEx neuron $i=1,\dots,N$ in the network is given by the system of coupled equations
\cite{destexhe09}
\begin{eqnarray}
C \frac{d V_{i}}{d t}  & = & - g_L (V_{i} - E_L) + g_L {\Delta}_T  \exp 
\left(\frac{V_{i} - V_T}{{\Delta}_T} \right) \label{eqadex1}\\
& & - \frac{1}{S}\left(w_{i} +  \sum_{j=1}^N g_{ij} (V_i-E_{{j}}) + \Gamma_i\right),\nonumber
\\
\tau_w \frac{d w_i}{d t}  & = &  a (V_i - E_L) - w_i,\label{eqadex2}
\end{eqnarray}
where $V_i$ and $w_i$ are, respectively, the membrane potential and adaptation
current of neuron $i$, $g_{ij}$ the synaptic conductance of the synapse from
neuron $j$ to neuron $i$, and $\Gamma_i$ the external perturbation applied
to neuron $i$. The synaptic conductance $g_{ij}$ has exponential decay with synaptic time-constant 
$\tau_s$. The parameter values in Eqs. (\ref{eqadex1}) and (\ref{eqadex2}) are given in Table \ref{table1}. 
These values have been chosen so that the system can reproduce the spiking characteristics of 
RS (excitatory) and FS (inhibitory) neurons observed in experiments with real neurons \cite{destexhe09}.

When the membrane potential of neuron $i$ is above a threshold potential ($V_i(t)>V_{\rm thres}=-30$ mV), 
the neuron is assumed to generate a spike and the following update conditions are applied
\begin{eqnarray}
V_i & \to & V_r  =  -60 {\, \rm mV}, \\ 
w_i & \to & w_i+b,\\
g_{ji} & \to & g_{ji} +g_s,
\end{eqnarray}
where $V_r$ is the reset potential. Parameters $b$ and $g_s$ have
different values depending on whether the neuron is excitatory or inhibitory: For excitatory neurons, 
$b=0.01$ nA and $g_s=g_{\rm ex}$, and for inhibitory neurons, $b=0$ and $g_s=g_{\rm in}$.
Updates have synaptic delays of $1.5$ ms and $0.8$ ms for excitatory and inhibitory synapses,
respectively. After the update, $g_{ji}$ decays exponentially with a fixed
time-constant $\tau_s$ ($5$ ms for excitatory and $10$ ms for inhibitory
synapses \cite{wang11}). We define the relative inhibitory conductance $g$ by
\begin{equation}\nonumber
g=g_{\rm in}/g_{\rm ex}
\end{equation}
as the parameter we will use in the investigation of network
dynamics. In each simulation, we apply external stimuli $\Gamma$ to $5\%$ of
the $N$ neurons (randomly chosen) for $50$ ms to initiate network activity, and
then stop the external stimuli to observe the activity triggered, which can be
persistent (SSA) or transient. For each neuron $i$, the external stimulus
$\Gamma_i$ has the same characteristics: it consists of excitatory current
pulses with synaptic conductances that rise instantaneously to 0.01 $\mu$S and
decay exponentially afterwards with a decay time of 5 ms, generated by a homogeneous 
Poisson process with rate 400 Hz.

All numerical simulations were implemented in C and the ordinary differential equations 
were integrated using the fourth order Runge-Kutta method with a fixed time step of $h=0.01$ ms.

\begin{table}[ht]
\centering
\caption{Parameter values used in Eqs. (\ref{eqadex1}) and (\ref{eqadex2}). Values for excitatory 
neurons are indicated by $^{\bullet}$ and for inhibitory by $^{\star}$. The ranges, where applicable, 
are indicated by square brackets. These values have been chosen so that the system of Eqs. 
(\ref{eqadex1}) and (\ref{eqadex2}) reproduces the spiking characteristics of RS (excitatory) and FS 
(inhibitory) neurons observed in experiments with real neurons \cite{destexhe09}.}
\label{table1}
\begin{tabular}{ccc}\hline \hline
Parameter                      & Symbol            &Value \\ 
\hline
\hline
Membrane capacitance              &$C$                &  1 $\mu{\rm F/cm}^2$ \\
Resting leak conductance          &$g_L$              & 0.05 ${\rm mS/cm}^2$ \\
Resting potential                 & $E_L$             & -60 mV \\
Slope factor                      & $\Delta_T$        & 2.5 mV \\
Spike threshold                   & $V_T$             & -50 mV \\
Membrane area                     & $S$               & 20,000 $\mu{\rm m}^2$ \\
Refractory time period            & $t_{r}$            & 2.5 ms \\
Adaptation intensity              & $a$               & 0.001 $\mu$S \\
Adaptation time constant          & $\tau_w$          & 600 ms \\
Integration time step             & $h$               & 0.01 ms \\
Time                              & $t$               & [0,10] s\\
Reversal potential                & $E_j$             & 0 mV $^{\bullet}$ \\
		                  &                   & -80 mV $^{\star}$ \\
Synaptic time constant            & $\tau_s$          & 5 ms $^{\bullet}$\\
		                  &                   & 10 ms $^{\star}$  \\
Synaptic conductance              & $g_s = g_{\rm ex}$  & [0,12] nS $^{\bullet}$ \\
		                  & $g_s = g_{\rm in}$  & [0,240] nS $^{\star}$ \\
Synaptic delay                    & $d$               & 1.5 ms $^{\bullet}$\\
		                  &                   & 0.8 ms $^{\star}$ \\
\hline
\hline
\end{tabular}
\end{table}

%%%%%%%%%%%%%%%%%%%%%%%%%%%%%%%%%%%%%%%
%%%%%%%%%%%%%%%%%%%%%%%%%%%%%%%%%%%%%%%

\section{Quantities for the study of self-sustained activity in numerical simulations and details of electrophysiological experiments}\label{dmee_section}

In this work, we use the following mathematical quantities to study SSA based either on time intervals or specific spike timings.

\subsection{Coefficient of variation}

The coefficient of variation is defined based on time intervals and exploits the
interspike interval (ISI) where the $m$th interval is defined as the
difference between two consecutive spike-timings $t^{m}_{i}$ and $t^{m+1}_{i}$ of
neuron $i$, namely ${\rm ISI_i}=t^{m+1}_{i}-t^{m}_{i}>0$. The coefficient of
variation of the $i$th neuron is then given by
\begin{equation}
\label{eqCV}
{\rm CV}_i=\frac{\sigma_{{\rm ISI}_i}}{\overline{{\rm ISI}_i}},
\end{equation}
where $\sigma_{{\rm ISI}_i}$ is the standard deviation of the interspike intervals of neuron $i$.

\subsection{Time-varying and mean network firing rates}

We define the spike-train of neuron $i$ as the sum of delta functions
\cite{gabbiani98}
\begin{equation}
\label{eqspktrain}
x_i(t) = \sum_{[t_i^m]}\delta(t - t_i^m),
\end{equation}
where $[t_i^m]$ is the set of all spike-timings of neuron $i$ for $t\in[0,T]$ and $T$ the final integration time.
Based on Eq. (\ref{eqspktrain}), we calculate the mean firing rate of neuron
$i$ over the time interval $[0,T]$ as
\begin{equation}
\label{eqfi}
\bar{F}_i = \frac{1}{T} \int_{0}^{T} x_i(t) dt.
\end{equation}
For all neurons in the network, we define the time-varying network firing rate
(in Hz) in intervals of $\psi=1$ ms as
\begin{equation}
\label{eqF}
F(t)=\frac{1}{\psi N}\sum_{i=1}^{N}\left(\int_{t}^{t+\psi}\delta(t-t_i^m)dt\right),
\end{equation}
and the mean network firing rate over the time interval $[0,T]$ as
\begin{equation}
\label{eqmeanF}
\bar{F} = \frac{1}{\langle{\rm ISI}\rangle},
\end{equation}
where $\langle{\rm ISI}\rangle$ is the mean ISI of the network given by
\begin{equation}
\label{eqmeanISI}
\langle{\rm ISI}\rangle=\frac{1}{N}\sum_{i=1}^{N}\langle{\rm ISI}_i\rangle.
\end{equation}

\subsection{Power spectrum, Fano factors and mean firing rate} \label{sub_sec_power}

Based on the definition of spike-trains $x_i(t)$ in Eq. (\ref{eqspktrain}), we define the power spectrum of neuron $i$ as
\begin{equation}
\label{eqSi}
S_i^{xx}(f) = \frac{\langle \tilde{x}_i(f) \tilde{x}_i^*(f)\rangle}{T},
\end{equation}
where $\langle \cdot \rangle$ indicates ensemble average, $T$ is the final integration time, $\tilde{x}_i(f)$ is the Fourier transform of neuron
$i$, given by
\begin{equation}
\tilde{x}_i(f) = \int\limits_0^T e^{2 \pi i f t} x_i(t)dt,
\end{equation}
and $\tilde{x}_i^*(f)$ is the complex conjugate of $\tilde{x}_i(f)$. The power
spectrum of a set of $M$ ($M\leq N$) neurons, $\bar{S}^{xx}(f)$ is then defined
as the average power spectrum of the $M$ neurons
\begin{equation}
\label{eqS}
\bar{S}^{xx}(f) = \frac{1}{M}\sum\limits_{i=1}^M{S}^{xx}_i(f).
\end{equation}
To obtain consistent results over different simulations (see Fig.
\ref{fig5}), we have chosen $M=5 \times 10^4$, independently of $N$ and $K$. In this work, we will use two quantities related to $\bar{S}^{xx}$ to describe
spike-train characteristics \cite{lindner09,grun2010,wieland15,pena18}.  

The first quantity is the Fano factor
$FF= \langle (n - \langle n \rangle)^2 \rangle / \langle n \rangle = \langle (\Delta n)^2 \rangle / \langle n \rangle$, 
which is defined as the
ratio of the variance of $n$ to the mean of the averaged spike count of the $M$ neurons over
the time window $[0,T]$, namely
\begin{equation}
n= \frac{1}{M}  \sum_{i=1}^{M} \int_0^T x_i(t)dt.\nonumber
\end{equation}
Its relation to $\bar{S}^{xx}$ is meaningful in the vanishing frequency limit of
$\bar{S}^{xx}(f)$, i.e. $\lim\limits _{f\to 0}\bar{S}^{xx}(f)= \bar{F} \cdot FF$.
The Fano factor, $FF$, is a standard measure of neural variability ($FF=1$ corresponds to a Poisson process) and is related to the CV of the ISIs of the $M$ neurons
\cite{cox66} by $\lim\limits _{f\to 0}\bar{S}^{xx}(f)= \bar{F} \cdot {\rm CV}^2
\left( 1 + 2 \sum_{k=1}^{\infty} r_k \right)$, where $r_k$ is the serial
correlation coefficient between ISIs that are lagged by $k$.

The second quantity is the mean firing rate of the $M$ neurons, $\bar{F}_M$ (defined as in Eq. (\ref{eqmeanF}) but dividing by $M$), which is related to
$\bar{S}^{xx}(f)$ by
\begin{equation}
 \bar{F}_M=\lim\limits _{f\to \infty}\bar{S}^{xx}(f).\label{ifl}
\end{equation}

\subsection{Synaptic input}

The instantaneous synaptic conductance of neuron $i$ due to all excitatory and
inhibitory synapses arriving to it, is denoted by
\begin{equation}
\label{eqGi}
G_i(t)=\sum_{j \in \mathfrak{E}} g_{ij}(t)+\sum_{j \in \mathfrak{I}}g_{ij}(t)=
G_{{\rm ex},i}(t)+G_{{\rm in},i}(t),
\end{equation}
where $\mathfrak{E}$ and $\mathfrak{I}$ are the sets of all excitatory and
inhibitory neurons in the network, respectively.

The instantaneous synaptic input to neuron $i$ is then defined by
\begin{eqnarray}
\label{eqIsynt}
I_{{\rm syn},i}(t) & = & \sum_{j \in \mathfrak{E}} g_{ij}(E_j - V_i) +
\sum_{j \in \mathfrak{I}} g_{ij}(E_j - V_i)\nonumber \\
& = & I_{{\rm syn},i}^{\rm E}(t) + I_{{\rm syn},i}^{\rm I}(t),
\end{eqnarray}
and the mean synaptic input $\bar{I}_{\rm syn}$ over all neurons in the network and over the simulation time in $[0, T]$ by  
\begin{equation}
\label{eqIsyn}
\bar{I}_{\rm syn} =  \frac{1}{N}\sum_{i=1}^{N} \frac{1}{T} \int_{t=0}^{T}
\left( I_{{\rm syn},i}^{\rm E}(t) + I_{{\rm syn},i}^{\rm I}(t) \right) dt .
\end{equation}

\subsection{Mean network decay-time}

Finally, for a given neuron $i$, we define the time of its last spike as the
maximum time in its spike train,
\begin{equation}
\label{eqtilast}
t_i^{\rm last} = \max{\{t_i^m\}}.
\end{equation}

The network decay-time (DT) is then defined by the maximum of the last spikes of all neurons in the network during the simulation time, i.e. the maximum $t_i^{\rm last}$ of all neurons in the network,
\begin{equation}
\label{eqDT}
{\rm DT} = \max{\{t_i^{\rm last},\; i = 1, \ldots, N\}}.
\end{equation}

Finally, we define the mean network decay-time ${\rm \overline{DT}}$ as the mean DT over $10^3$ simulations with different random initial conditions.

\subsection{Details of electrophysiological experiments}

\subsubsection{Animals}

Electrophysiological experiments were conducted using male Wistar rats with
20-25 postnatal days. All animals were kept in an animal facility in a 12:12h
light-dark cycle at a temperature of 23$^\circ$C $\pm$ 2$^\circ$ with free access
to food and water.  All procedures were approved by the Institutional Animal
Care Committee of the Institute of Biomedical Sciences, University of S\~ao
Paulo (CEUA ICB/USP n. 090, fls. 1$^\circ$).

\subsubsection{Preparation of brain slices}

After animals were deeply anesthetised through isoflurane inhalation
(A\-Errane; Baxter Pharmaceuticals), they were decapitated and the brain was
quickly removed and submerged in cooled (0$^\circ$C) oxygenated (5\%
CO$_2$-95\% O$_2$) cutting solution (in mM): 206 sucrose, 25 NaHCO$_3$, 2.5
KCl, 10 MgSO$_4$, 1.25 NaH$_2$PO$_4$, 0.5 CaCl$_2$, and 11 D-glucose. After
removing the cerebellum, brain hemispheres were separated by a single sagittal
cut. Both brain hemispheres were trimmed up and glued in a metal platform and
sectioned using a vibratome (Leica - VT1200). 350-400 $\mu$m brain slices were
obtained by advancing the vibratome blade from anterior-posterior orientation.
Slices were rapidly transferred to a holding chamber containing artificial
cerebrospinal fluid (ACSF; in mM): 125 NaCl, 25 NaHCO$_3$, 3 KCl, 1.25
NaH$_2$PO$_4$, 1 MgCl$_2$, 2 CaCl$_2$, and 25 D-glucose. Slices were kept
oxygenated at room temperature (20-25$^\circ$) for at least one hour before
proceeding with electrophysiological recordings. 

\subsubsection{Electrophysiological recordings}

Brain slices containing the hippocampal formation were placed in a
sub\-mersion-type recording chamber upon a modified microscope stage and
maintained at 30$^\circ$C with constant perfusion of oxygenated ACSF (5\%
CO$_2$-95\% O$_2$). Whole-cell recordings were made from neurons located in the pyramidal layer of CA1. Recording pipettes were fabricated from borosilicate
glass (Garner Glass) with input resistances of about 4-6 M$\Omega$ and were
filled with intracellular solution (in mM): 135 K-gluconate, 7 NaCl, 10 HEPES,
2 Na2ATP, 0.3 Na3GTP, 2 MgCl$_2$; at a pH of 7.3 obtained with KOH and
osmolality of 290 mOsm. All experiments were performed using a visualised slice
setup under a differential interference contrast-equipped Nikon Eclipse E600FN
microscope. Recordings were made using a Multiclamp 700B amplifier and pClamp
software (Axon Instruments). Only recordings from cells that presented
spontaneous activity with membrane potentials lower than $-60$ mV, access
resistance lower than 20 M$\Omega$, and input resistance higher than 100
M$\Omega$ and lower than 1000 M$\Omega$, were included in our data. We injected depolarising currents to identify regular, tonic, or bursting spike patterns and neural spontaneous activity was assessed by 10 min of continuous
recordings in current clamp mode.

%%%%%%%%%%%%%%%%%%%%%%%%%%%%%%%%%%%%%%%
%%%%%%%%%%%%%%%%%%%%%%%%%%%%%%%%%%%%%%%

\section{Self-sustained activity}\label{ssa_section}

\subsection{Self-sustained activity in electrophysiological experiments}

SSA assessed by electrophysiological recordings in physiological brain states
is characterised by irregular neural spiking, normally with low average
frequency where rates follow a long-tailed distribution across neurons
\cite{litwin12}. Interestingly, some brain regions are able to produce
spontaneous network activity after a slicing procedure, including hippocampal
sharp waves \cite{maier03,Giannopoulos13,bazelot16}. Once hippocampal slices
present SSA manifested by spontaneous activity, we use CA1 neurons whole-cell
recordings to demonstrate the possible variability of firing rate patterns
observed in the brain. Our intracellular recordings show a high variability of
spontaneous activity patterns including low neural firing rates and
small-bursts activity in distinct recorded neurons.

\begin{figure}[ht]
\centering
\includegraphics[height=8cm,width=13cm]{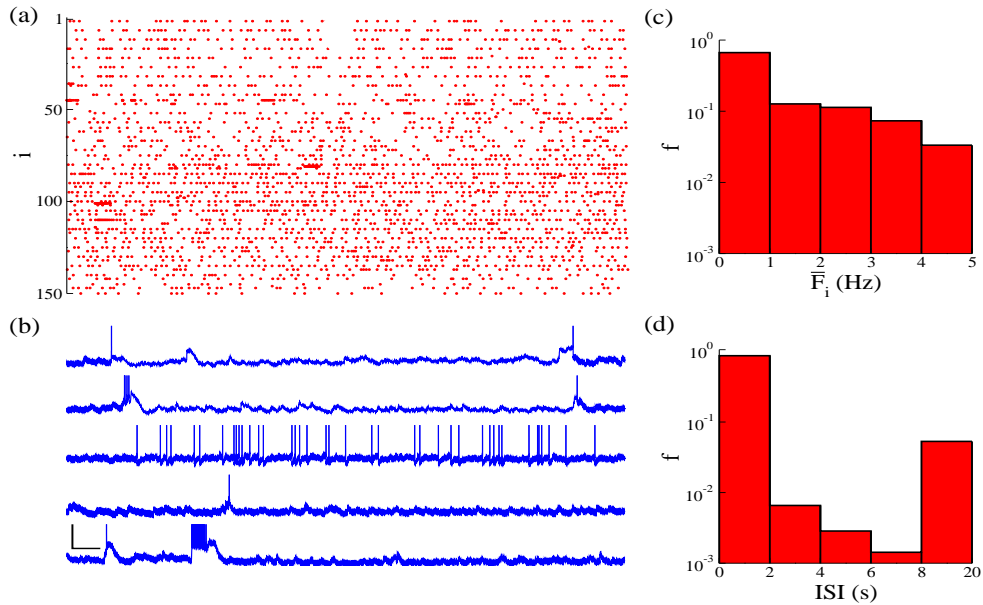}
\caption{Results from electrophysiological experimental data. (a) Raster plot of 150 traces. (b) Twenty seconds of recordings
are shown (horizontal black bar denotes $1$ s and vertical black bar denotes
$50$ mV). (c) Mean firing rate and (d) interspike interval (ISI)
distributions. Note that the vertical axes in (c) and (d) are in logarithmic scale and that the label $f$ represents the fraction of values in the bins in the horizontal axes.}
\label{fig1}
\end{figure}

Figure \ref{fig1} shows a representation of traces obtained by whole-cell patch
clamp from CA1 neurons during a period of $20$ s. The raster plot of the 150
traces is shown in Fig. \ref{fig1}(a), and five traces of membrane potential
are shown in Fig. \ref{fig1}(b) for the same time window. The mean firing rate
and interspike interval (ISI) distributions are plotted in Figs. \ref{fig1}(c)
and \ref{fig1}(d), respectively. Recorded neurons present distinct firing
patterns, including very low firing rate and small bursts of spikes, as shown
in Fig. \ref{fig1}(b). In the whole record, the mean firing rate over all
neurons is approximately equal to $1.172$ Hz. Similar firing patterns, with
high variability and low neural firing rates, are found in different
recordings in the hippocampus and other cortices of the rat brain during slow-wave
sleep \cite{mizuseki13}, as well as in recordings from human middle temporal gyrus
during sleep \cite{peyrache12}.

\subsection{Self-sustained activity in numerical simulations}

Here, we use the results from numerical simulations of the a random network model of $N$ adaptive 
exponential integrate-and-fire neurons \cite{Brette2005} (Eqs. (\ref{eqadex1}) and (\ref{eqadex2})). 
We focus on reproducing three aspects of the previous subsection (electrophysiological experimental data): 
i) mean firing rates around 1 Hz, ii) small-bursts activity in distinct neurons 
(Fig. \ref{fig1}(b)) and iii) data with variability in firing rates where most of the neurons have 
frequencies between 0 and 1 Hz (Fig. \ref{fig1}(c)).

The random network model of $N$ adaptive, exponential integrate-and-fire neurons
\cite{Brette2005} we use here is composed of 80\% excitatory and 20\% inhibitory neurons (i.e. a ratio of 4:1) following \cite{noback05}. The synaptic conductance shows an exponential decay with synaptic delays of $1.5$ ms and $0.8$ ms for excitatory and inhibitory synapses, respectively. In each simulation, we apply external stimuli $\Gamma$ to $5\%$ of the $N$ neurons (randomly chosen) for $50$ ms to initiate network activity. Then, we stop the external stimulation to observe the activity triggered, which can be persistent (i.e. SSA) or transient.

\begin{figure}[ht]
\centering
\includegraphics[height=8cm,width=13cm]{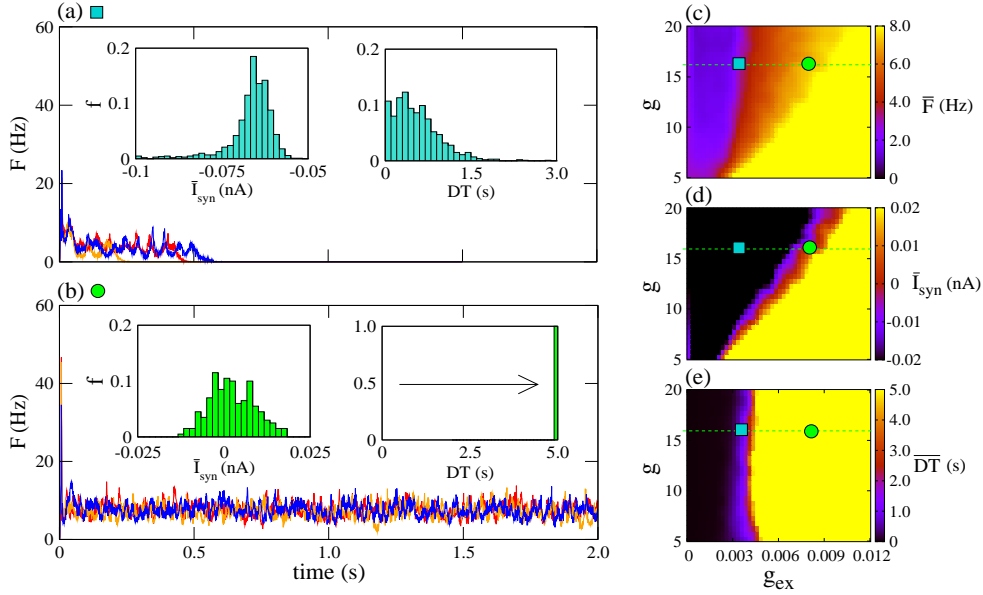}
\caption{Results from the numerically simulated data. Time-dependent network firing rate ($F$) in two distinct regimes: (a)
non-SSA and (b) SSA. Each case corresponds to a different simulation. The two
neural network models use the same set of parameters except that the excitatory
conductance is $g_{\rm ex}=0.0035$ $\mu$S in (a) and $g_{\rm ex}=0.008$ $\mu$S in (b). The insets show the probability distribution of the mean synaptic inputs ($\bar{I}_{\rm syn}$) over $10^3$ simulations and decay-times (DT) in each case. (c) Mean network firing rate ($\bar{F}$), (d) mean synaptic input
($\bar{I}_{\rm syn}$) and (e) Mean decay-time (${\rm \overline{DT}}$) as a function of excitatory
($g_{\rm ex}$) and relative inhibitory synaptic coupling ($g$). The neural
network has $N=10^4$ neurons and the connection probability is $p=0.02$. In (a) and (b), we fixed $g=16$. The turquoise squares correspond to the parameters in (a) and the green circles to those in (b). Note that the label $f$ in the insets in (a) and (b) represents the fraction of values in the bins in the horizontal axes.}
\label{fig2}
\end{figure}

To reproduce SSA firing patterns with low neural firing rates, we analyse the parameter space $g_{\rm ex} \times g$ for a neural, random, network of $N=10^4$ AdEx neurons with connection probability $p=0.02$. We focus on an area of the parameter space where there is a balanced regime of excitation and inhibition. In this area, we do not observe SSA for $g_{\rm ex}<0.004$ $\mu$S, i.e. in the weak coupling region. Examples of time-dependent network firing
rates in this region are shown in Fig. \ref{fig2}(a), for $g_{\rm ex}=0.0035$
$\mu$S and $g=16$. The initial stimulus applied in the first $50$ ms, generates
short-lived activity in the network. The mean synaptic input, $\bar{I}_{\rm syn}$, is negative (left-hand inset in Fig.
\ref{fig2}(a)) and the network decay-time DT is always less than $3$ s
(right-hand inset in Fig. \ref{fig2}(a)). Examples of time-dependent network
firing rates in the region of SSA are shown in Fig. \ref{fig2}(b),
corresponding to $g_{\rm ex}=0.008$ $\mu$S and $g=16$. The mean
synaptic input is approximately balanced ($\bar{I}_{\rm syn} \approx$ 0) and the
decay-time is higher than the maximum time used in our simulations (i.e. DT $\ge 5$ s).

For $N=10^4$ neurons, the lowest mean network firing rates are approximately
$2$ Hz. The region where these rates occur is shown in purple in Fig.
\ref{fig2}(c). A particular case for $g_{\rm ex}<0.0035$ $\mu$S is indicated by a
turquoise square, where the activity is not self-sustained. For the random
network and parameter space considered, the region with SSA is roughly
determined by $g_{\rm ex}>0.004$ $\mu$S (indicated in yellow in Fig.
\ref{fig2}(e)). The lowest mean network firing rate of an SSA state is around
$4$ Hz (red region in Fig. \ref{fig2}(c)). Within this region, one can identify
the region of excitation/inhibition balance by considering the region with
$I_{\rm syn}\approx 0$ (see Eq. (\ref{eqIsyn})). This is shown in red in Fig.
\ref{fig2}(d). The black and yellow regions in Fig. \ref{fig2}(d) correspond to
slightly predominant inhibitory and excitatory mean synaptic input,
respectively. An SSA case with $\bar{F} \approx 4$ Hz and excitation/inhibition
balance is indicated by the green circle in Figs. \ref{fig2}(c)-(e) and
corresponds to $g_{\rm ex}=0.008$ $\mu$S and $g=16$. 

\begin{figure}[ht]
\centering
\includegraphics[height=7.5cm,width=13.5cm]{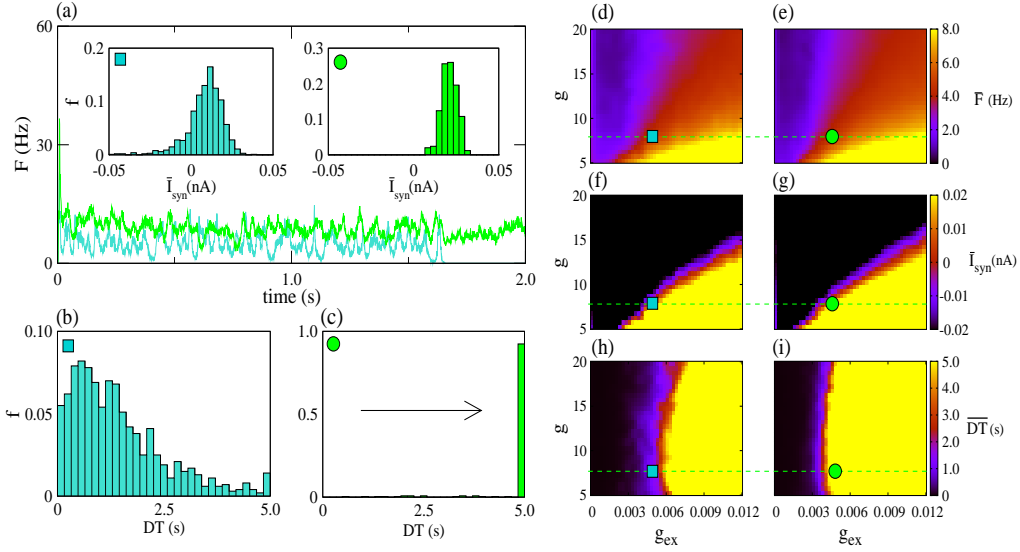}
\caption{Results from the numerically simulated data. (a) Time-dependent network firing rate ($F$) for $N=10^4$ (turquoise
line) and $N=2 \times 10^4$ (green line). The insets show the probability
distribution of the mean synaptic input
($\bar{I}_{\rm syn}$) over $10^3$ simulations, for each case. (b) and (c) show the decay-time (DT) distribution (also for $10^3$ simulations) for $N=10^4$ and $N=2 \times 10^4$, respectively (colours follow the same convention as in (a)). We considered the excitatory conductance
$g_{\rm ex}=0.005$ $\mu$S, mean node-degree $K=400$, and relative
inhibitory synaptic coupling $g=8$. The mean network firing rates (d) and (e),
mean synaptic inputs (f) and (g), and mean decay-times (h) and (i) are shown as a
function of $g_{\rm ex}$ and $g$ for $N=10^4$ (left) and $N=2 \times 10^4$
(right), respectively. The turquoise squares correspond to $N=10^4$ and
$p=0.04$ and the green circles to $N=2 \times 10^4$ and $p=0.02$. Note that the label $f$ in the insets in (a) and (b) represents the fraction of values in the bins in the horizontal axes.}
\label{fig3}
\end{figure}

To study the effect of the size of the network on SSA, we considered two 
networks of different sizes with the same mean degree. The first has $N=10^4$ and
$p=0.04$ and the second $N=2 \times 10^4$ and $p=0.02$, both having $K=400$.
The parameters ($g_{\rm ex}=0.005$ $\mu$S and $g=8$) are the same for the two
networks, and put them close to the balanced state. In Fig. \ref{fig3}(a), we
show the firing rate evolution of the two networks. For $N=10^4$ (turquoise
line), the activity decays before $2$ s, and for $N=2 \times 10^4$ (green line)
SSA is observed. The distributions of $\bar{I}_{\rm syn}$ for the two cases (see
the insets in Fig. \ref{fig3}(a)) have similar (positive) average values but
the one for $N=10^4$ is broader and left-skewed. The decay-time distribution
(Figs. \ref{fig3}(b), (c)) clearly shows that the networks with $N=2 \times
10^4$ have SSA, while the ones with $N=10^4$ are predominantly short-lived. A
comparison of networks with the two sizes in the $g_{\rm ex}\times g$ parameter
space is shown in Figs. \ref{fig3}(d)-(i). The mean network firing rate and
mean synaptic input display similar behaviour in the parameter space for the two
network sizes (Figs. \ref{fig3}(d)-(g)). However, this similarity is not seen
in the diagram for ${\rm \overline{DT}}$ (Figs. \ref{fig3}(h)-(i)). The region 
corresponding to SSA (yellow) is larger for $N=2 \times 10^4$ than for $N=10^4$.
Moreover, the shape of this region for $N=2 \times 10^4$ discloses almost
absent sensitivity of SSA duration to the relative inhibitory synaptic
conductance $g$. On the other hand, for $N=10^4$, the SSA lifetime is sensitive
to $g$ for $0.006$ $\mu {\rm S} \lesssim g_{\rm ex} \lesssim 0.008$ $ \mu
{\rm S}$, and only for strong coupling ($g_{\rm ex} \gtrsim 0.008$ $ \mu
{\rm S}$), it becomes insensitive to $g$. In Fig. \ref{fig3}, we observe 
large decay time DT for high frequencies.

\begin{figure}[ht!]
\centering
\includegraphics[height=8cm,width=11cm]{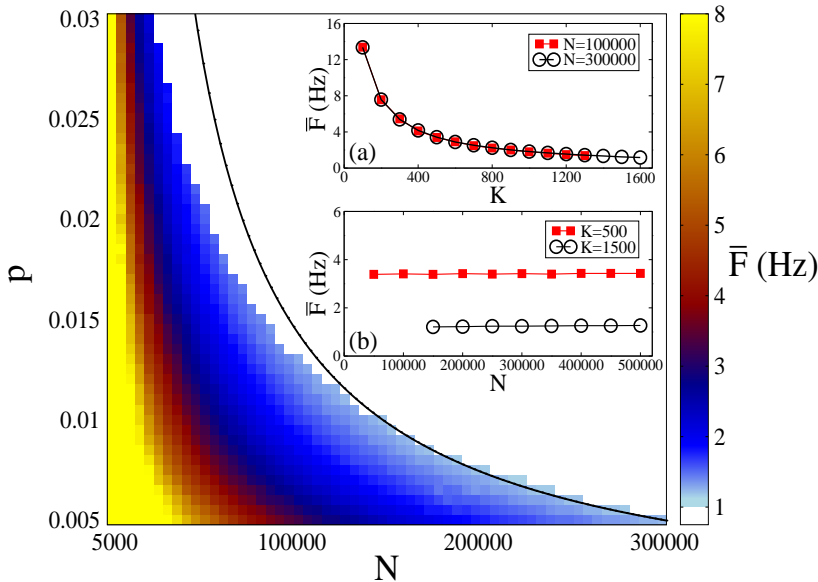}
\caption{Results from the numerically simulated data. Each point in the coloured part of the plot gives $\bar{F}$ calculated for a simulation with the corresponding parameter values of $N$ and $p$. The white area corresponds to non-SSA states. The black line represents networks with mean node-degree $K=1500$ (according to Eq. (\ref{eqpK})). Upper inset: mean network firing rate $\bar{F}$ as a function of $K$, keeping $N$ fixed, where the red squares and black circles represent $N = 10^5$ and $N = 3 \times 10^5$ neurons, respectively. Lower inset: mean network firing rate $\bar{F}$ as a function of $N$, showing independence of $\bar{F}$ on $N$ for constant connection degree $K$. The curves represent two selected $K$ values: $K=500$ (red squares) and $K=1500$ (black circles). Note that we have used $g_{\rm ex}=0.008$ $\mu$S and $g=16$ for all simulations in this figure.}
\label{fig4}
\end{figure}

Next, we investigate the influence of $N$, $p$ and $K$
(see Eq. (\ref{eqpK})) on the network firing rate of SSA states. We fix $g_{\rm ex}=0.008$ $\mu$S and $g=16$ to keep the dynamics around an
excitatory/inhibitory balance. Figure \ref{fig4} shows the mean network firing
rate of SSA states (colour scale) in the parameter space $N\times p$. The white area in the parameter space corresponds to non-SSA states. We can see that the mean network firing rate of SSA states depends on both $N$ and $p$. In
particular, low firing rate SSA states appear when the network size $N$
increases. The black solid line in Fig. \ref{fig4} represents networks with
mean node-degree $K=1500$, in which case there are SSA states for $N \geq 1.5
\times 10^5$ neurons. The inset in Fig. \ref{fig4}(a) shows the dependence of
the mean network firing rate $\bar{F}$ of SSA states on the mean connection
node-degree $K$ for constant $N$. We can observe that lower rates are obtained for increased $K$. However, very low firing rates ($\approx$ $1$ Hz) are present only for very large network sizes. In the inset in Fig. \ref{fig4}(b), we can
observe that the network size does not alter the mean network firing rate when
$K$ is fixed. Therefore, large $K$ plays an important role in the
occurrence of low network firing rates, and for such low firing rates, large-size networks are necessary to support SSA states.

\begin{figure}[ht]
\centering
\includegraphics[height=8cm,width=13cm]{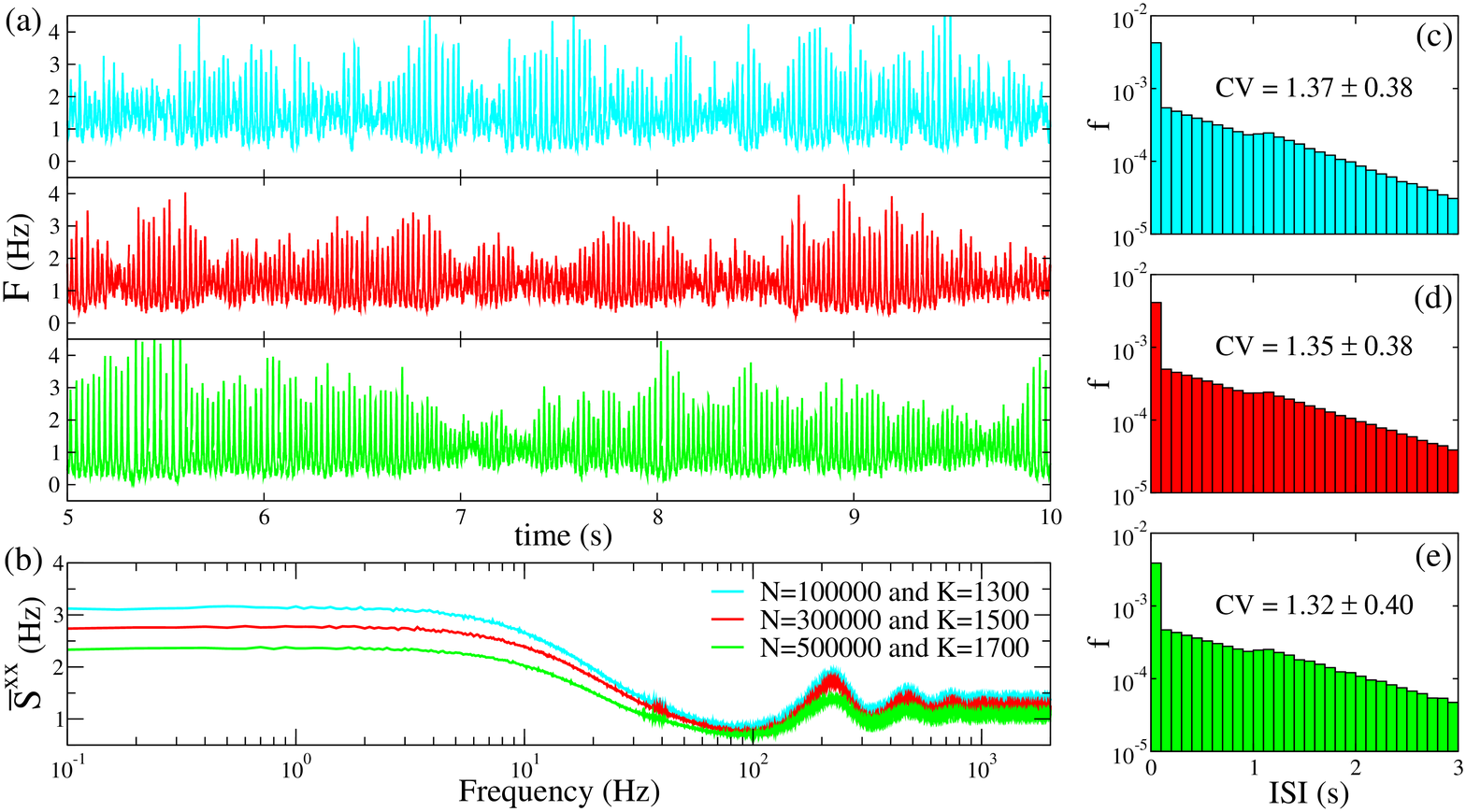}
\caption{Results from the numerically simulated data. (a) From top to bottom, we show $F$ as $N$ and $K$ increase (for the $N$ and $K$ values, see legend in (b)). (b) Power spectra averaged over $5 \times 10^4$ neurons in the network and $10$ s of simulation. Note that the horizontal axis is logarithmic and the vertical linear. (c)-(e) ISI distributions for the same three networks with the vertical axes in logarithmic and horizontal in linear scales. The values of the corresponding CVs are shown in the plots. In the three cases, the parameter values used are $g_{\rm ex}=0.008$ $\mu$S and $g=16$. Note that the vertical axes in (c), (d) and (e) are in logarithmic scale and that the label $f$ represents the fraction of values in the bins in the horizontal axes.}
\label{fig5}
\end{figure}

Figure \ref{fig5} shows results for three cases of network sizes and mean node-degrees $K$, namely for $N = 10^5$ and $K = 1300$, for $N=3\times 10^5$ and $K = 1500$ and for $N = 5 \times 10^5$ and $K = 1700$. The time-varying network firing rate is non-periodic and its mean value decreases as both $N$ and $K$ increase (Fig. \ref{fig5}(a)). In Figs. \ref{fig5}(c)-(e), we observe that most ISIs are distributed in the interval $[0,2]$ s. These results agree with the experimental results in Fig. \ref{fig1}(d), where more than 60\% of ISIs are in a similar interval. In the three cases, the spiking variability, characterised by the ISI distribution, is very well described by a Poisson distribution (Figs. \ref{fig5} (c)-(e)). This is characteristic of irregular neural firing \cite{maimon09,mochizuki16,Lundqvist2010,ostojic14}. Moreover, the CV of the ISIs slowly converges to the one of a Poisson process ($=1$) as both $N$ and $K$ increase.

As discussed in Sec. \ref{dmee_section}, the power spectrum of a set
of $M$ neurons is related to different quantities that
characterise network firing. In our numerical simulations, we can observe these relations by taking the limits of the power spectrum as $f\to \infty$ or $f\to 0$. The mean firing rate $\bar{F}_M$ is associated with the infinity frequency limit, namely $\lim _{f \rightarrow \infty} \bar{S}^{x x}(f)=\bar{F}_M$
 (see also Eq. (\ref{ifl})). In Fig. \ref{fig5}(b), one can
see that these mean firing rates are very low ($1$ Hz
$\lesssim \bar{F}_M \lesssim 1.5$ Hz) and that the smaller rates occur for larger $N$ and $K$. On the other hand, irregularity is associated with
vanishing frequency, namely $\lim\limits_{f\to 0}\bar{S}^{xx}(f) = \bar{F}.FF$, where $FF$ is the Fano factor (see Subsec. \ref{sub_sec_power}). Assuming approximately the same $\bar{F}$ for the three networks in Fig. \ref{fig5}(b), say
$\bar{F} \approx 1.3$, the $f\to 0$ limit of $\bar{S}^{xx}(f)$ shows that
$FF>1$ for all of them. This suggests a process more irregular than the Poisson
process \cite{vanvreeswijk04}, e.g. bursting \cite{mochizuki16}. Moreover, as
the network size $N$ increases, $FF$ decreases, indicating tendency to converge
to a Poisson process and is linked to a standard irregularity measure and CV.
We see that in the limit of very small frequencies, the value of the power
spectra decreases as both $K$ and $N$ increases, confirming the behaviour
observed by CV, i.e. the spiking times are becoming more regular. The power
spectra display slow fluctuations, which can be explained by the low neural
firing rates and bursting spiking patterns (as discussed below). Slow power
spectrum fluctuations are also features of spontaneous activity in cortical
networks \cite{mantini07,wieland15,mastrogiuseppe17}.

\begin{figure}[ht]
\centering
\includegraphics[height=7.6cm,width=13cm]{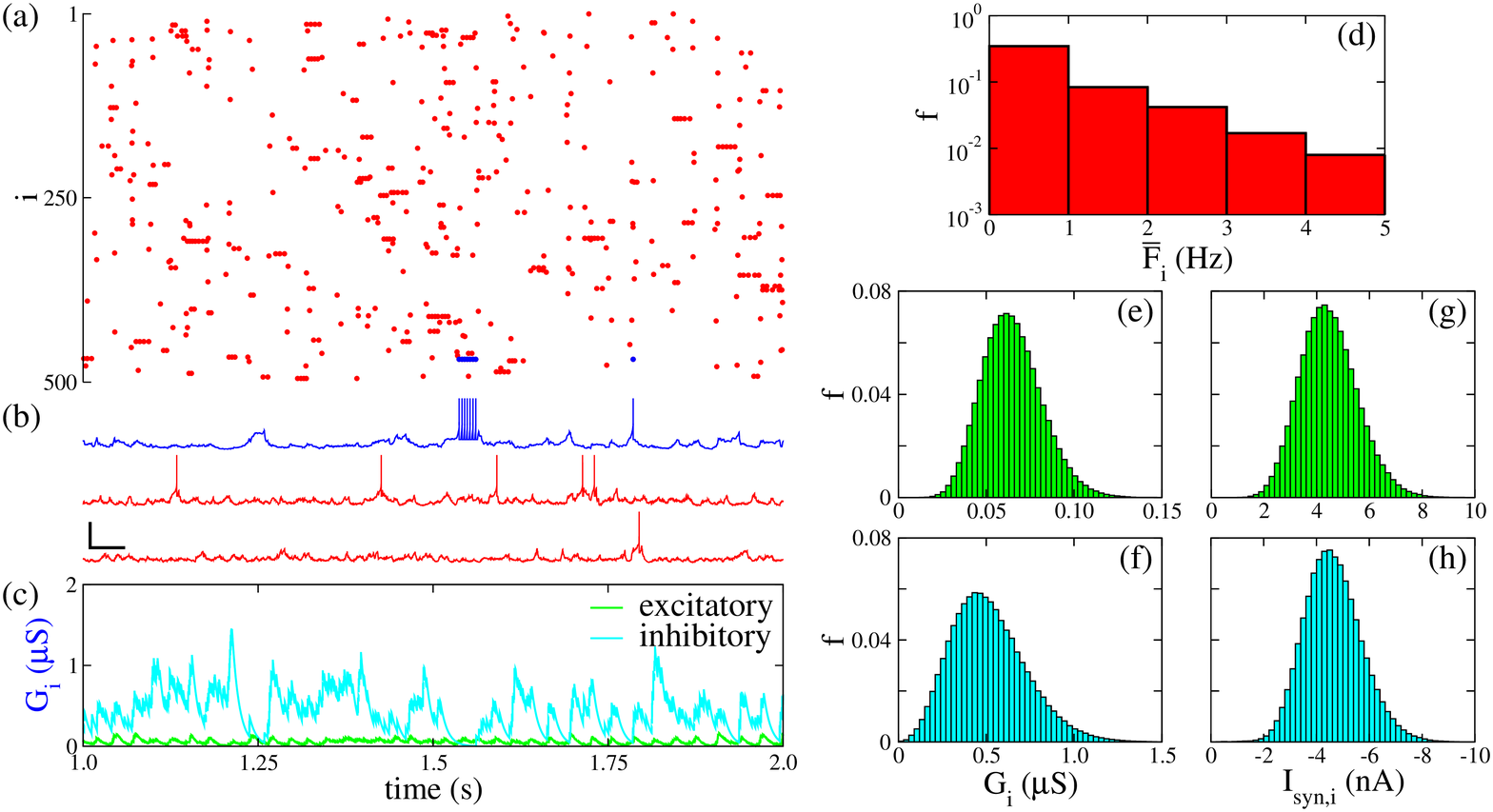}
\caption{Results from the numerically simulated data. All plots refer to a single simulation of a network with $N=5 \times 10^5$, $K=1700$, $g_{\rm ex}=0.008$ $\mu$S and $g=16$. (a) Raster plot of 2 s of simulation time (only 500 randomly chosen neurons from the network are shown). (b) membrane potential of three randomly chosen neurons from the network (the horizontal black bar denotes 0.2 s and the vertical black bar denotes 50 mV). (c) Time-varying synaptic conductances ($G_{{\rm ex},i}(t)$ in green and $G_{{\rm in},i}(t)$ in cyan) of neurons represented by the blue line in (b). (d) distribution of mean firing rates $\bar{F}_i$ (Eq. (\ref{eqfi})) of all neurons in the network. Distribution of excitatory ($G_{{\rm ex},i}$) (e) and inhibitory ($G_{{\rm in},i}$) (f) synaptic conductances of all neurons at the end of the simulation. Distribution of excitatory ($I_{{\rm syn},i}^{\rm E}$) (g) and inhibitory ($I_{{\rm syn},i}^{\rm I}$) (h) synaptic inputs of all network neurons at the end of the simulation. Note that the vertical axis in (d) is in logarithmic scale and that the label $f$ in (d), (e), (g), (f) and (h) represents the fraction of values in the bins in the horizontal axes.}
\label{fig6}
\end{figure}

In Fig. \ref{fig6}, we show characteristics of the SSA firing patterns with low
rate exhibited by the network (a particular case with $N=5 \times 10^5$ and
$K=1700$ is shown as a representative example). The patterns are characterised by sparse and
non-synchronous activity (Fig. \ref{fig6}(a)), akin to what has been termed the
heterogeneous variant \cite{ostojic14} of the asynchronous irregular (AI)
regime \cite{brunel00,vogels05b}. In the homogeneous AI regime, all neurons
fire with the same mean rate, but in the heterogeneous AI regime, the mean
firing rates fluctuate in time and across neurons. In some cases, neurons can
even exhibit bursting periods (as can be seen in Fig. \ref{fig6}(b)). Neurons
have low firing rates, with a right-skewed distribution that peaks around 1 Hz
(Fig. \ref{fig6}(d)). Irregularity is confirmed by the CV of ISIs and power
spectra analysis (Fig. \ref{fig5}). The similarity between the firing regime
observed in our simulations and the heterogeneous AI state can be further
verified by a comparison of the respective spike-train power spectra. Spectra
for networks in heterogeneous AI have been calculated elsewhere \cite{pena18}
and display slow fluctuations as shown here (Fig. \ref{fig5}(b)).

The distributions of the excitatory ($G_{\rm ex}$) and inhibitory ($G_{\rm in}$) synaptic conductances over neurons (panels (e) and (f) in Fig. \ref{fig6}) have nearly symmetric shapes, with the $G_{\rm in}$ distribution slightly right skewed, in agreement with experimental evidence \cite{rudolph07}, even though the conductance values per se are higher than in the experimental recordings. The distributions of excitatory ($I_{{\rm syn}}^{\rm E}$) and inhibitory ($I_{{\rm syn}}^{\rm I}$) synaptic inputs are nearly identical and vary over nearly the same range of absolute values, though the upper end of the range is slightly higher for the inhibitory synaptic inputs (see panels (g) and (h) in Fig. \ref{fig6}). This is a hallmark of a balanced state, and the neural spikes occur due to the large synaptic conductance variability, which allows for the necessary conditions that can support the appearance of SSA.

Finally, it is worthwhile the fact that in Figs. \ref{fig6}(a) (raster plot) and \ref{fig6}(b) (membrane potential), there are several striking similarities, namely: non-synchronous activity and, the coexistence of spike and bursting neural activity. More importantly though, qualitatively similar activities and behaviours are present in the data obtained from hippocampal slices (electrophysiological experiments) as demonstrated in Fig. \ref{fig1}(a)-(b) and in the results from the numerical simulations data. In both experimental and numerically simulated data, we observe high variability in neural firing rates where most of the neurons fire with frequencies in the interval $[0,1]$ Hz, with similar distribution of mean firing rates in the interval $[1,5]$ Hz (compare Fig. \ref{fig1}(c) and Fig. \ref{fig6}(d)).

%%%%%%%%%%%%%%%%%%%%%%%%%%%%%%%%%
%%%%%%%%%%%%%%%%%%%%%%%%%%%%%%%%%

\section{Conclusions and discussion}

In this paper, motivated by self-sustained activity observed in the brain in the absence of external stimuli, we sought to study necessary conditions for which irregular and low-frequency self-sustained dynamics emerge in models of neural networks with random connectivity. We build neural network models of $80\%$ excitatory and $20\%$ inhibitory neurons randomly connected and mathematically described by the adaptive, exponential integrate-and-fire model. This model mimics the behaviour of biological neurons, exhibiting spiking and bursting patterns of activity. We studied network features by varying: (i) the balance between excitation and inhibition in the system and (ii) the topological characteristics of the random network by means of varying the mean node-degree. Results were obtained by running a large number of numerical simulations to compute firing rates and related quantities, and compared them with results from our electrophysiological experiments using whole-cell patch clamp from CA1 rat neurons.

Our numerical results allowed us to observe neural activity with slow fluctuations in the absence of external perturbation. We found that the patterns of low firing-rate self-sustained activity is asynchronous and irregular as depicted in raster plots. We shown that the irregular spikes with low firing rate depend on the mean node degrees of the neurons, and that low-rate self-sustained activity occurs for a tight balance between inhibition/excitation and large network sizes. We found that for fixed mean connection degree, the mean network firing rate does not change with the network size. In an excitation/inhibition balanced random network, if the network size is fixed, the mean network firing rate decreases when the mean node-degree increases. However, there is a maximum mean node-degree value for which it is possible to maintain self-sustained activity states, and this value is proportional to the network size. Therefore, large networks are necessary to give rise to self-sustained activity states when the mean node-degree is large. Moreover, the occurrence of spikes is due to the synaptic-conductance variability and the inhibitory synaptic input is slightly higher than the excitatory.

More importantly, we found that our results from the numerical simulations resemble those obtained from our experimental recordings and analysis, such as irregular firing with firing rates of about 1 Hz. In addition, we have demonstrated that our model is able to reproduce the coexistence of spike and bursting neurons and the distribution of mean firing rates between 0 and 5 Hz obtained by our electrophysiological experiments using whole-cell patch clamp from CA1 rat neurons.

Concluding, we have been able to demonstrate the existence of low neural firing rate self-sustained activity in excitation/inhibition balanced random networks of adaptive, exponential integrate-and-fire neurons. This phenomenon is characterised by irregular neural oscillations and shows that low frequency self-sustained activity can be found in large balanced random networks, a result that is qualitatively similar to those obtained from our electrophysiological experiments using whole-cell patch clamp from CA1 rat neurons.

%%%%%%%%%%%%%%%%%%%%%%%%%%%%%%%%%
%%%%%%%%%%%%%%%%%%%%%%%%%%%%%%%%%

\section*{Acknowledgements}
We wish to acknowledge the support from the following grants and sche\-mes:
CAPES (88881.120309/2016-01), CNPq (154705/2016-0, 306251/2014-0, 311467/2014-8, 432429/2016-6), FAPEAL, International Visiting Fellowships Scheme of the University of Essex, and FAPESP (2011/19296-1, 2013/ 07699-0, 2013/25667-8, 2015/50122-0, 2015/ 07311-7, 2016/23398-8, 2017/ 13502-5, 2017/18977-1, 2017/20920-8, 2017/26439-0, and 2015/50122-0).

\end{document}